\overfullrule=0pt %
PARAMETRIC EPICYCLIC RESONANCE IN BLACK HOLE DISKS: QPOs IN MICRO-QUASARS
\bigskip
W. Klu\'zniak $^1$ and M. A. Abramowicz$^2$
\bigskip
\par\noindent
$^1$Institute of Astronomy, Zielona G\'ora University, Lubuska 2,
65-365 Zielona G\'ora, Poland
\par\noindent
$^2$Department of Astronomy and Astrophysics, Chalmers University,
412-96 G\"oteborg, Sweden
\bigskip
\centerline{Abstract}
Non-linear acoustic coupling of modes
in accretion disks allows a parametric resonance between epicyclic motions.
In black hole disks, such a resonance first occurs, and is strongest,
 when the radial and vertical
epicyclic frequencies are approximately in a 2:3 ratio, in agreement
with the 300 and 450 Hz frequencies reported in GRO J1655-40
and  the 184 and 276 Hz frequencies reported in XTE J1550-564.
The first overtone in the same resonance is in a 5:3 ratio
with the fundamental, in agreement with the 69.2 and 41.5 Hz frequencies
reported in GRS 1915+105. The narrow width and the frequency stability of
the corresponding QPOs follow from general properties of parametric resonance.
\bigskip
Introduction
\bigskip

We give an explanation for the presence in black hole X-ray sources of the
observed stable frequencies, manifest as narrow-width
QPOs (quasi-periodic oscillations in the light curve). 
These oscillations arise as a result of parametric resonance
in the accretion disk, and the actually observed rational ratios of
frequencies correspond to that resonance which is strongest {\it a priori}.

Different modes in the disk are coupled through terms involving
derivatives of enthalpy. Oscillations of pressure in one mode lead to
harmonic variations of the eigenfrequency in another mode.
In the limit of small pressure corrections, particular ratios of
the epicyclic frequencies correspond to the condition for parametric
resonance, leading to exponential growth of one of the modes.
This is a specifc mechanism through which the previously suggested
(Klu\'zniak and Abramowicz 2001a,b; Abramowicz and Klu\'zniak 2001)
resonant origin of high frequency QPOs in neutron stars and black holes
can come about. 
For a discussion of the relation of this phenomenon to ``quantization
of orbits'' see Abramowicz et al 2002. 
Further, we point out that coherent localized structures
may form in the accretion disk.

Strohmayer (2001a) reports 300 and 450 Hz QPOs in GRO J1655-40,
and we have already pointed out that these are in a 2:3 ratio
(Abramowicz and Klu\'zniak 2001, Klu\'zniak and Abramowicz 2001a,b).
Strohmayer 2001b reports 
$41.5\pm0.4\,$Hz and $69.2\pm0.15\,$Hz in GRS 1915+105,
 giving 0.599 as the frequency
ratio, rather close to 3:5=0.6 (if taken at face value).
These ratios seemed to us to be indicative of a resonant phenomenon,
and the recent discovery of another pair of commensurate frequencies
(in a 2:3 ratio),
184 and 276 Hz in XTE J1550-564 (Remillard et al. 2002; see also
Miller et al. 2001),
greatly strengthens the case for resonance.

\bigskip
Localized structures in the disk
\bigskip

We find, that ``geostrophic'' flow is a solution of the equations of
disk hydrodynamics and this would correspond to a localized elliptic
flow pattern, counter-rotating with
the radial epicyclic frequency in the co-moving fluid frame. 
 We point out that close to the inner edge of
a thin accretion disk around black holes (and neutron stars),
the vorticity becomes very small, and goes to zero where the radial
epicyclic frequency vanishes, at the marginally stable orbit.
It is, therefore, not difficult to excite a vorticity
perturbation of large {\it relative} amplitude in that region.
This is in contrast with the Newtonian case of a $1/r$ potential,
for which Spiegel and coworkers find that an initial amplitude 
above a finite threshold is necessary for a vorticity perturbation
to persist in the disk  (Bracco et al., 1998).

Such localized structures may execute ``up and down'' motions
away from the disk plane, and it follows from the equations of
hydrodynamics (acceleration is equal to the gradient
of the gravitational potential plus the gradient of enthalpy),
that such motion is harmonic with frequency
$$\omega_2^2=\Omega_z^2 +\epsilon,$$
while the corresponding harmonic motion in the radial direction
occurs at frequency
$$\omega_1^2=\Omega_r^2 + \gamma^2\epsilon,$$
where $\Omega_r$ is the radial epicyclic frequency,
$\Omega_z$ the vertical epicyclic frequency,
 $\epsilon$, assumed to be $<<1$, is the pressure correction
and $\gamma^2$ is the ratio of vertical to radial extent of the disturbance.
The equation for $\omega_1$ can be thought of as the high, or low, frequency
limit of the general dispersion relation of diskoseismology
(Wagoner 1999, Kato 2000),
$\epsilon=c_s^2k^2$ corresponds to the (``acoustic'') $p$-mode,
while a negative value of $\epsilon$ corresponds to the (``internal gravity'')
$g$-mode. We share the view stressed by diskoseismologists
(e.g., Wagoner, Silbergleit, and Ortega 2001) that the very stability
of the observed frequencies, while the system changes its luminosity
(and presumably its accretion rate), implies that pressure forces are
a minor perturbation. 

\bigskip
Parametric resonance
\bigskip

We are trying to explain how QPOs may arise from minor fluctuations
in radial motion. There is a mechanism which leads to large amplitudes
of oscillatory motion, given small perturbations. It is parametric
resonance. In the case under discussion, radial motions at frequency
$\omega_1$ will disturb the pressure at the same frequency, i.e., will
induce harmonic variations in the value of $\epsilon$. Then, 
the eigenfrequency
of vertical oscillations, $\omega_2$, will itself undergo small
harmonic variations in its value, occurring at frequency
$\omega_1\approx\omega_r$. This is precisely a description of
parametric resonance in an oscillator with the equation of motion
$d^2z/dt^2+\omega_2^2[1+h\cos(\omega_1t)]z=0$. Resonance
occurs when $\omega_1=2\omega_2/n$, with $n$ integer
(Landau and Lifschitz 1973).
For  $\omega_1$ in a small, and decreasing with $n$, range about
$2\omega_2/n$, exponential growth of the amplitude of oscillations
occurs until non-linear saturation. If a dissipation term is present,
growth occurs only if a certain threshold value is exceeded by the amplitude,
$h$, of variations in the value of the oscillator parameter.

The relatively large amplitude vertical oscillation can then in turn
amplify radial oscillations through the usual resonance mechanism
in driven oscillators.

\bigskip
Frequency ratios
\bigskip

Under the assumption (justified above) that $\omega_1\approx\Omega_r$,
and $\omega_2\approx\Omega_z$, we note that parametric resonance
in black hole accretion disks occurs only for $n\ge3$. This is simply because
in the Kerr metric,  $\Omega_r<\Omega_z$, always, and the condition
for parametric resonance is $\omega_1=2\omega_2/n$.
The fastest growing resonant mode in black hole disks occurs
when $\Omega_r=2\Omega_z/3$, i.e.,
 for the lowest possible value of $n$,
 when the two epicyclic frequencies
are in a 2:3 ratio, because in general
the lower the value of $n$, the faster the growth of amplitude.
It is remarkable that in at least two black hole sources
this is the ratio of the observed frequencies.
We interpret these observed frequencies as the fundamental frequency
of vertical oscillations excited through parametric resonance by
oscillations at the radial epicyclic frequency, at 2/3 the value of the
fundamental.

In non-linear resonances, combination frequencies appear in addition
to the fundamental. In the case of a parametric 2:3 resonance,
the first ``combination'' overtone appears at 5/3 the value of the fundamental.
The 69.2 Hz frequency reported   in GRS 1915+105 is indeed 5/3 the
value of the also reported 41.5 Hz. This is highly suggestive.
We note that under this interpretation, if the mass of the source
is indeed as given by Greiner et al. (2001) , i.e.,
it is no more than 18 or $20M_\odot$, this
source would have to be a Schwarzschild black hole, or the spin
parameter would have to have small negative values, i.e., the disk
would be counter-rotating. In any case the magnitude of $j$ would be
much less than for the other two sources, GRO J1655-40 and XTE  J1550-564.

A subharmonic at the frequency 1/3 of the fundamental would also be
expected, and this has also been reported, for XTE  J1550-564
(Remillard 2002).

\bigskip
Conclusions
\bigskip

Parametric resonance between epicyclic frequencies in a thin-disk
accretion flow in the Kerr metric can lead to the excitation
of finite amplitude oscillations and this can
account for the appearance  of commensurate ``stable''frequencies
in the observed sources. In addition to the strongest oscillations
in the 2:3 ratio, subharmonics and overtones are predicted, yielding
a sequence of primes 1:2:3:5, all of which have been observed (albeit
in various sources).

\bigskip
References
\bigskip
\parindent=-0.1truein

Abramowicz, M.A., et al. 2002, Class. Quant. Grav. in press, gr-qc/0202020

Abramowicz, M.A., Klu\'zniak, W. 2001, Astron. Astrophys. 374, L19,
astro-ph/0105077

Bracco, A., Provenzale, A., Spiegel, E.A., Yecko, P. 1998,
in Theory of Black Hole Accretion 
\par~~~
Disks, M.A.~Abramowicz, G.B. Bj\"ornsson,
J.E. Pringle, eds., Cambridge university Press, p. 254

Greiner, J., Cuby, J.G., McCaughrean, M.J. 2001, Nature 414, 522

Kato, S. 2000, Pub. Astron. Soc. Japan, 52, 1133

Kato, S. 2001, Pub. Astron. Soc. Japan, 53, L37

Klu\'zniak, W., Abramowicz, M.A. 2001a, astro-ph/0105057,

Klu\'zniak, W., Abramowicz, M.A. 2001b, Acta Phys. Polon. B 32, 3605

Landau, L.D., Lifschitz, E.M. 1973, Classical Mechanics, (Pergamon)

Miller, J. et al. 2001, ApJ 563, 928

Remillard, R. et al. 2002, astro-ph/0202305

Strohmayer, T.E. 2001a, Astrophys. J. 552, L49

Strohmayer, T.E. 2001b, Astrophys. J. 554, L37

Silbergleit, A. S., Wagoner, R. V., Ortega-Rodríguez, M. 2001, ApJ, 548, 335

Wagoner, R.V. 1999, Phys. Rep. 311, 259
\bye